# Assessing the reliability and validity ranges of magnetic characterization methods

Mohammad Reza Zamani Kouhpanji[1, 2] and Bethanie J H Stadler[1, 3, *]

[1]Department of Electrical and Computer Engineering, [2]Department of Biomedical Engineering, [3]Department of Chemical Engineering and Materials Science, University of Minnesota Twin Cities, USA.

**Abstract**

Evaluating the interaction fields of magnetic nanowires (MNWs) is of utmost importance for advancing their functionality in diverse applications including spintronic devices and nanomedicine. In recent years, several quantitative methods have been proposed and become inevitable tools to quantify the interaction fields and decouple their effects from the coercivity. However, the uncertainty of the attained results arose countless open questions leading to discrepancies among the literature. Here, we employ our novel experimental method, named the projection method, to resolve these discrepancies. Using a comparative analysis of the four most commonly used methods (hysteresis loops method, remanence curves method, first-order reversal curve method, and projection method), we unambiguously explicate the reliability and validity limits of these methods to elucidate the origin of the discrepancies. We show that the remanence curves method must solely be used for quantifying the interaction fields if they are considerably weaker compared to the coercivity. Furthermore, we show that both remanence curves method and first-order reversal curve method fail to fully decouple the interaction fields' effects from the coercivity, similar to the hysteresis loops method.

Miniaturization of magnetic materials resulted to emerge of a variety of magnetic nanowires (MNWs), such as magnetic/non-magnetic multi-segmented[1,2] and/or diameter modulated[3–5], with superb quantum efficiency that has never be realized in their bulk states. As a result, the MNWs have been proposed in diverse applications including magnetic recording media[6], magnetic refrigeration[7,8], nanosensors and nanoactuators[9,10], microwave nanodevices[11–13], spintronic nanodevices[14–17], as well as biology and nanomedicine[18–23]. In all these applications, the intrinsic properties of the MNWs must be determined precisely because they play a critical role in the functionality of the whole device[24–28]. Due to the quantum entanglement effects, the interaction fields substantially impact the intrinsic properties of MNWs especially in assemblies and/or arrays. These effects cause a significant misunderstanding of the intrinsic properties and functionality of the MNWs. For example, the interaction fields can cause artificial magnetic frustration[29,30], artificial shift of the magnetic resonance frequency[31,32] and energy barrier[33], broadening of the intrinsic switching field[34,35], reduction of the heating efficiency[22], superparamagnetic behavior[36], to name a few. Consequently, it is crucial to reliably characterize the interaction fields to determine the intrinsic properties of MNWs.

---

[*] Corresponding author, Email: stadler (at) umn (dot) edu.

Both interaction fields and coercivity depend on the dimensions and geometry of the MNWs arrays that makes discriminating them extremely difficult[25]. In this direction, numerous theoretical and experimental approaches have been introduced. Theoretically, micromagnetic simulations have been employed to underline the nature of the interaction fields and their effects on the coercivity to accurately determine intrinsic properties. Due to the computational limitations, these simulations are typically limited to short-range MNWs with perfect geometry and homogenous properties that do not meet experiments[37–40]. The experimental methods, on the other hand, do not suffer from these ideal assumptions as they are directly conducted on real MNWs. However, their application owes understanding two critical facts: 1) the range of the reliability and validity of the magnetic measurements and 2) the unambiguous analysis for determining the interaction fields and coercivity.

Based on the representation of the results, the experimental methods can be categories into two groups, in which the first group qualitatively describes the interaction fields and coercivity while the second group quantitatively describes them. A few examples for the former group are recoil curves method[41,42], Henkel and the $\delta M$ method[43–45]. And, a few examples for the latter group are the minor loop shifts[46], energy barrier shifts[33], ferromagnetic resonance[31,32], and $\delta H$ method[47,48]. The major drawback of the quantitative methods is that they do not directly measure the interaction fields and coercivity. As a result, several attempts have been done over the decades to advance the qualitative methods for quantitative analysis of these parameters. For example, the remanence curves have been analyzed with respect to the Stoner-Wohlfarth model to extract the interaction fields[25,49]. Another example is the integration of the first-order reversal curves (FORC) distributions over field axes to quantify the interaction fields and coercivity[50–53].

In this work, we characterize the interaction fields ($H_{int}$) and coercivity ($H_c$) of MNWs arrays using the most popular magnetic measurements, the hysteresis loops method, remanence curves method, FORC method, and our novel projection method to assess their reliability and validity limits. To do so, eight different types of the nickel MNWs were fabricated using the template-assisted electrodeposition technique, details are given in the SI. The MNWs are categorized into two categories based on their templates, random arrangement with low porosity and ordered arrangement with high porosity, where both categories have the same range of diameters. The randomly arranged MNWs were achieved using track-etched polycarbonate membranes with an average diameter (porosity) of 30nm (0.5%), 50nm (1%), 100nm (2%), and 200nm (12%). The ordered arranged MNWs were achieved using anodic aluminum oxide (AAO) membranes with an average diameter (porosity) of 20nm (12%), 80nm (15%), 120nm (17%), and 200nm (20%). We chose these two categories to have MNWs with relatively similar $H_c$ while significantly different levels of $H_{int}$ in addition to be able to investigate the capability of the aforementioned methods for characterizing highly randomized and highly ordered arrangements. Experimentally, the magnetic measurements (hysteresis loops, FORC, and remanence curves measurements) were conducted and analyzed using the standard protocols. We also measured the magnetic response of the MNWs using our well-established projection method protocols as we introduced[26,54,55]. The raw data for all magnetic measurements are given in the SI. Since the hysteresis loops method does not characterize the $H_{int}$ and its effects on the $H_c$, we consider it as the bottom-line to compare other methods demonstrating their reliability and validity limits.

Before quantifying the $H_{int}$ in the MNWs, we first qualitatively illustrate the $H_{int}$ across the different types of the MNWs. Regardless of the qualitative methods for describing the $H_{int}$, a simple and fast qualitative visualization of the $H_{int}$ can be achieved by analyzing the squareness of the MNWs hysteresis loop, defined as the ratio of the saturation remanence magnetization to the saturation magnetization. Both saturation remanence magnetization and saturation magnetization are determined from the upper branch of the hysteresis loop. Higher $H_{int}$ causes shearing of the hysteresis loop leading to a reduction of the squareness of the hysteresis loop. Another approach to qualitatively determine the $H_{int}$ is the reversibility fraction, defined as the ratio of reversible magnetization to the total magnetization of the MNWs arrays[26,56]. In contrast to the squareness criteria, the reversibility fraction increases as the $H_{int}$ increases. Figure 1 shows the squareness and reversibility fraction of the MNWs arrays. Qualitatively, as the porosity increases, the reversibility fraction increases while the squareness decreases for all MNWs types. As it could be expected, increasing the porosity qualitatively increases the $H_{int}$ regardless of the arrangement and/or the level of the porosity. Note that both squareness and reversibility fraction can be determined using the projection method with a single measurement while the hysteresis loops and FORC methods do not provide the reversibility fraction. Furthermore, neither of these parameters can be measured using the remanence curves method.

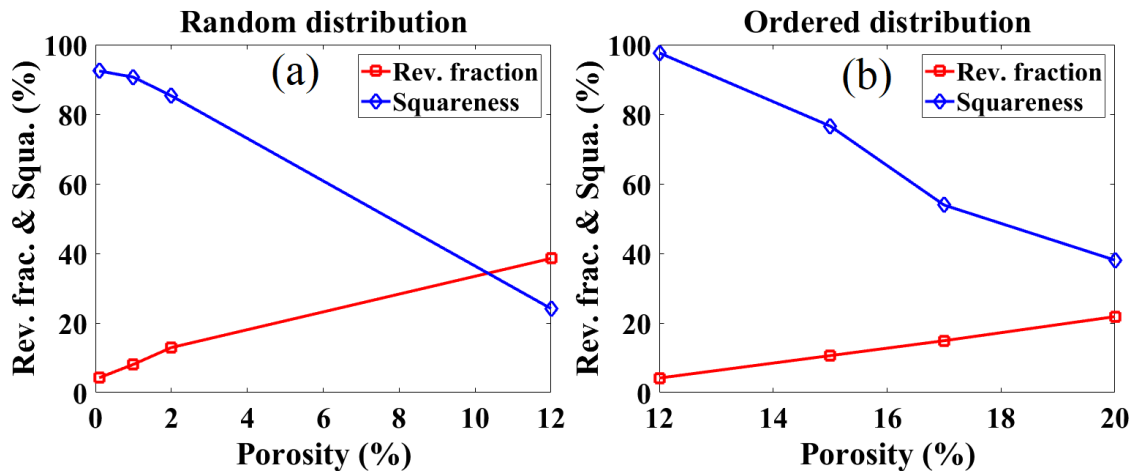

Figure 1: illustrating the porosity effects on the reversibility fraction and squareness of the hysteresis loop for (a) randomly distributed MNWs and (b) orderly distributed MNWs.

Figure 2 depicts the $H_{int}$ for the different types of MNWs arrays. Among all methods, the remanence curves method determines the $H_{int}$ much smaller than the other methods. Furthermore, this method surprisingly measures a reduction in the $H_{int}$ as the porosity increases even though it shows an increase in the $H_{int}$ for small porosities. Here, the remanence curves, isothermal remanence curve and DC demagnetization curve, were measured according to the standard protocols and analyzed according to[25]. This anomalous behavior of the remanence curves method is due to two facts. First, it employs the first-order mean-field theory to calculate the $H_{int}$ that considers the $H_{int}$ linearly proportional to the magnetization. It has been shown that the first-order mean-field theory is valid only for exchange decoupled MNWs with $H_{int}$ much smaller than $H_c$[49,57]. Second, the remanence curves method calculates the $H_{int}$ from the difference of fields, where the normalized isothermal remanence and the normalized DC demagnetization are 1/3[25]. If the $H_{int}$ is large, the remanence magnetizations keep reconfiguring the MNWs magnetization state, equivalently, reducing the $H_{int}$. This phenomena occur until a balance between the $H_{int}$ and the $H_c$ is reached. At this point, the $H_{int}$ is no longer strong enough to overcome the MNWs $H_c$. This means that the remanence curves method indeed does not measure the actually $H_{int}$ but it measures a reduced $H_{int}$ where the MNWs are magnetically stable. The situation becomes worse at very large $H_{int}$. Therefore, the $H_{int}$ calculated using the remanence curves method is much smaller than other values and it decreases as the porosity increases.

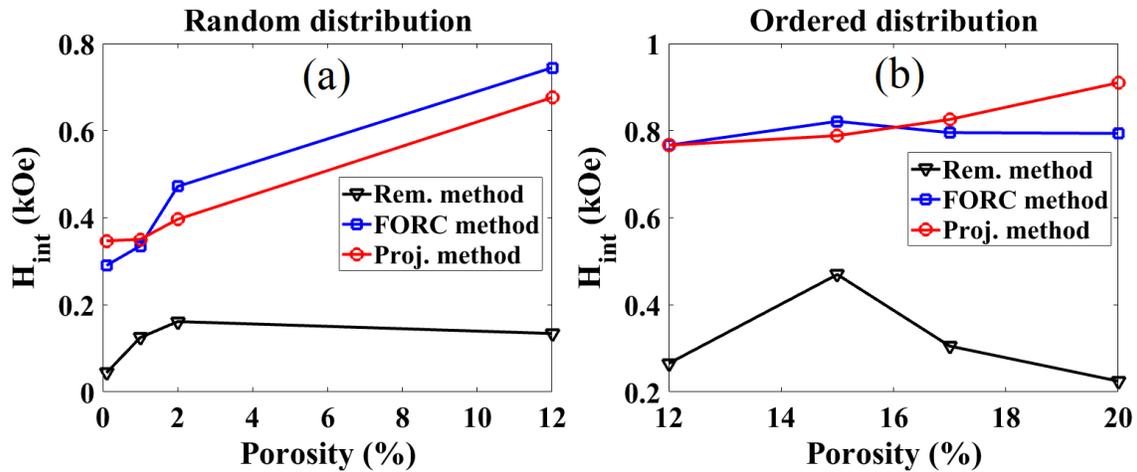

Figure 2: A quantitative analysis of the interaction field ($H_{int}$) calculated using different methods. In both subfigures, the "Rem." and "Proj." stand for "remanence curve" and "projection", respectively.

Figure 2 also shows the $H_{int}$ calculated using the FORC method and the projection method. Both methods calculate the $H_{int}$ much larger than the remanence curves method. The FORC and projection methods measure the $H_{int}$ fairly in the same ranges for both randomly and orderly distributed MNWs arrays. For both categories of the MNWs arrays, the projection method renders a trend that can be used to characterize the effects of exchange coupling, dipole fluctuation, and dipole-dipole coupling on the $H_{int}$. The FORC method, on the other hand, measures the $H_{int}$ fairly constant for the orderly distributed MNWs arrays even though the qualitative analysis, Figure 1, renders an increase in the $H_{int}$ in terms of the porosity. This is due to the destructive data analysis of the FORC method that can cause erasing the real features while adding artificial features[54,55,58,59]. Practically, the FORC method requires taking two derivatives that was shown to delete both irreversible magnetizations and the reversible magnetizations. Furthermore, taking two derivatives amplifies the measurement noises leading to spurious features in the FORC distributions. Smoothing can be applied to mitigate the noises up to a certain level but the smoothing effects are questionable as it makes the data interpretation more complex by drastically altering the FORC distributions[60–63].

Figure 3 depicts the $H_c$ of the MNWs arrays measured using all methods in addition to the $H_{int}$ in terms of diameters to facilitate the comparison. Practically, the hysteresis loop method does not determine the $H_{int}$, so its results for $H_c$ are contaminated by the $H_{int}$. Furthermore, since the remanence curves method relays on the first-order mean-field theory to determine the $H_{int}$ and its value is zero at H= $H_c$, the remanence curves technically does not determine the actual $H_c$[25,49] for interacting MNWs arrays. This fact is also supported by the $H_c$ values that it calculates as they are smaller and or similar to the $H_c$ values from the hysteresis loop. More interesting, the FORC method calculates the $H_c$ in the same range that could be attributed due to several reasons. First, as already mentioned, the ambiguous and complicated data processing of the FORC method miscalculate the $H_c$. Second, the FORC distributions do not exclusively show the probably of finding a MNW flipping down and up at the reversal field and applied field, respectively[37,40,54]. Indeed, they present the MNWs flipping up and down histogram where the MNWs with lower $H_c$ have the highest contributions on the FORC distributions. As a result, regardless of accumulation of the noises during the integral for finding the $H_c$, the MNWs with lower $H_c$, especially those bearing higher $H_{int}$, become dominant on the $H_c$ results[28]. Furthermore, it can be anticipated that the FORC method does not fully decouple the $H_{int}$ effects from the $H_c$. It is notable that the FORC method measures very large values for the standard deviation of the $H_c$, shown by $\delta H_c$ in Figure 3d. The $\delta H_c$ must be due to the non-uniformity of the MNWs[64,65]; however, the SEM images do not show a significant deviation in the diameters, see SI. Furthermore, larger $\delta H_c$ basically means broader switching fields that was shown to be due to the $H_{int}$[34,35]. Note the hysteresis loop and remanence curve method do not provide any information regarding the $\delta H_c$.

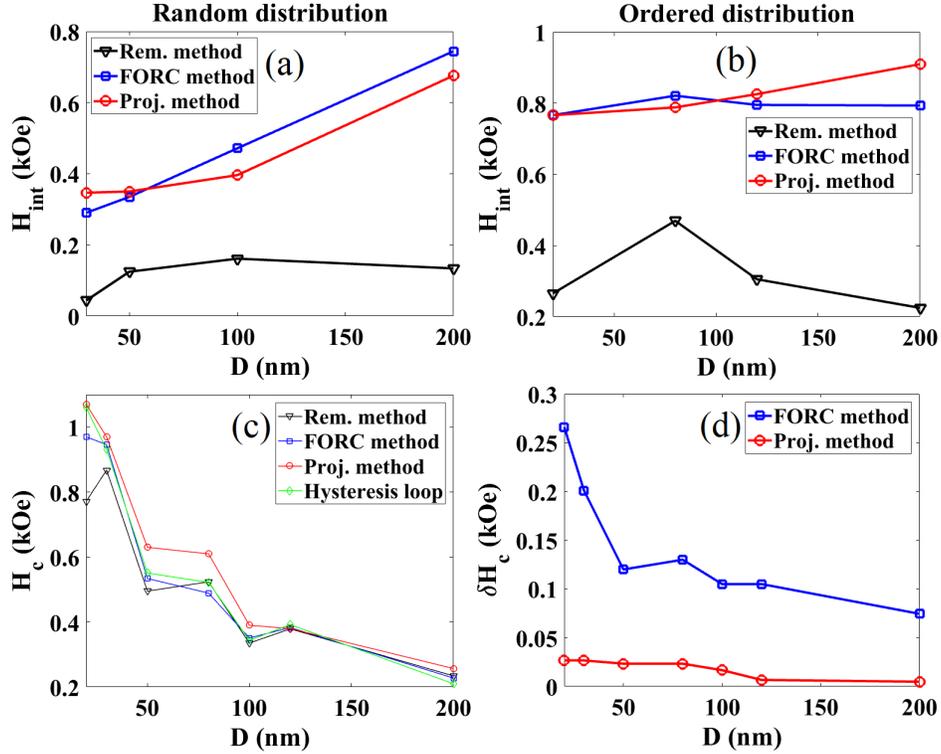

Figure 3: Plotting the interaction field ($H_{int}$) and coercivity ($H_c$) as a function of diameters.

In summary, we demonstrated the reliability and validity limits of the most popular current state of the art of the magnetic methods for quantifying the $H_{int}$ and $H_c$. Since the hysteresis loop method does not decouple the $H_{int}$ effects from the $H_c$, we considered it as the bottom-line to analysis other quantitative methods. The remanence curves method is only applicable to magnetically hard MNWs where the $H_{int}$ is significantly smaller than the $H_c$. Furthermore, since this method employs the first-order mean-field theory to calculate the $H_{int}$, it does not essentially discard the $H_{int}$ effects from the $H_c$. The FORC method, on the other hand, can be used for both magnetically soft and hard MNWs if the MNWs have a narrow $H_c$ distribution. Indeed, for the MNWs with a broad $H_c$ distribution, the FORC method determines the $H_c$ of the weakest MNWs because they have multitude contributions on the FORC distribution. Furthermore, regardless of the very time-consuming measurements and data analysis of the FORC method, this method does not fully decouple the effects of the $H_{int}$ from the $H_c$. Among all of the investigated quantitative method, the projection method exhibits a universality for measuring the $H_{int}$ and $H_c$ because it does not employ the first-order mean-field theory and it has a very straightforward analysis regardless of the type and arrangement of the MNWs.

**References**

[1] E. Berganza, M. Jaafar, C. Bran, M. Vázquez, and A. Asenjo, 1 (2017).

[2] R. Mendoza-reséndez, C. Luna, D. Görlitz, and K. Nielsch, **1047**, 1041 (2014).

[3] J.A. Fernandez-Roldan, R. Perez Del Real, C. Bran, M. Vazquez, and O. Chubykalo-Fesenko, Nanoscale **10**, 5923 (2018).

[4] C. Bran, E. Berganza, J.A. Fernandez-Roldan, E.M. Palmero, J. Meier, E. Calle, M. Jaafar, M. Foerster, L. Aballe, A. Fraile Rodriguez, R.P. Del Real, A. Asenjo, O. Chubykalo-Fesenko, and M. Vazquez, ACS Nano **12**, 5932 (2018).

[5] D.M. Burn, E. Arac, and D. Atkinson, Phys. Rev. B - Condens. Matter Mater. Phys. **88**, 1 (2013).

[6] J. Um, M.R. Zamani Kouhpanji, S. Liu, Z. Nemati, J. Kosel, and B. Stadler, Trans. Magn. - Conf. **56**, (2019).

[7] J.H. Belo, A.L. Pires, J.P. Araújo, and A.M. Pereira, J. Mater. Res. **34**, 134 (2019).

[8] D.I. Bradley, A.M. Guénault, D. Gunnarsson, R.P. Haley, S. Holt, A.T. Jones, Y.A. Pashkin, J. Penttilä, J.R. Prance, M. Prunnila, and L. Roschier, Sci. Rep. **7**, 1 (2017).

[9] P.D. McGary, L. Tan, J. Zou, B.J.H. Stadler, P.R. Downey, and A.B. Flatau, J. Appl. Phys. **99**, (2006).

[10] S.M. Reddy, J.J. Park, S.M. Na, M.M. Maqableh, A.B. Flatau, and B.J.H. Stadler, Adv. Funct. Mater. **21**, 4677 (2011).

[11] E. V Tartakovskaya, *Spin Waves and Electromagnetic Waves in Magnetic Nanowires* (Elsevier Ltd., 2015).

[12] C. Caloz and D.L. Sounas, in *IEEE MTT-S Int. Microw. Symp. Dig.* (IEEE, 2012), pp. 1–3.

[13] M. Sharma, B.K. Kuanr, M. Sharma, and A. Basu, in *J. Appl. Phys.* (2014), pp. 1–4.

[14] S.S.P. Parkin, M. Hayashi, and L. Thomas, Science (80-. ). **320**, 190 (2008).

[15] S. Parkin and S. Yang, Nat. Publ. Gr. **10**, 195 (2015).

[16] H. Mohammed, S. Al Risi, T.L. Jin, J. Kosel, S.N. Piramanayagam, and R. Sbiaa, Appl. Phys. Lett. **032402**, (2020).

[17] E.M. Palmero, M. Méndez, S. González, C. Bran, V. Vega, M. Vázquez, and V.M. Prida, Nano Res. **12**, 1547 (2019).

[18] Z. Nemati, J. Um, M.R. Zamani Kouhpanji, F. Zhou, T. Gage, D. Shore, K. Makielski, A. Donnelly, and J.A. Masa, ACS Appl. Nano Mater. **3**, 2058 (2020).

[19] A.P. Safronov, B.J.H. Stadler, J. Um, M.R. Zamani Kouhpanji, J.A. Masa, A.G. Galyas, and G. V Kurlyandskaya, Materials (Basel). **12**, 1 (2019).

[20] J.A. Masa, H. Khurshid, V. Sankar, Z. Nemati, M.H. Phan, E. Garayo, J.A. Garcia, and H. Srikanth, J. Appl. Phys. **113**, (2016).

[21] D.E. Shore, D. T, J. Modiano, M.K. Jenkins, and B.J. Stadler, Sci. Rep. **8**, 1 (2018).

[22] D.E. Shore, A. Ghemes, O. Dragos-Pinzaru, Z. Gao, Q. Shao, A. Sharma, J. Um, I. Tabakovic, J.C. Bischof, and B.J.H. Stadler, Nanoscale **11**, (2019).

[23] M.R. Zamani Kouhpanji, J. Um, and B.J.H. Stadler, ACS Appl. Nano Mater. (2020).

[24] J. Geshev, R.F. Lopes, J.L. Salazar Cuaila, L.L. Bianchi, and A. Harres, J. Magn. Magn. Mater. **500**, 166420 (2020).

[25] E. Araujo, J.M. Martínez-Huerta, L. Piraux, and A. Encinas, J. Supercond. Nov. Magn. **31**, 3981 (2018).

[26] M.R. Zamani Kouhpanji, P.B. Visscher, and B.J.H. Stadler, ArXiv **1**, 1 (2020).

[27] G. Tsoi and L. Wenger, (2014).

[28] C.I. Dobrotă and A. Stancu, J. Appl. Phys. **113**, (2013).

[29] H. Ge, Y. Xie, and Y. Chen, Phys. B Condens. Matter **577**, 411826 (2020).

[30] S.H. Skjærvø, C.H. Marrows, R.L. Stamps, and L.J. Heyderman, Nat. Rev. Phys. **2**, 13 (2020).

[31] J. De La Torre Medina, L. Piraux, J.M. Olais Govea, and A. Encinas, Phys. Rev. B - Condens. Matter Mater. Phys. **81**, 1 (2010).

[32] M. Demand and L. Piraux, **63**, 1 (2001).

[33] C. Moya, Ó. Iglesias, X. Batlle, and A. Labarta, J. Phys. Chem. C **119**, 24142 (2015).

[34] L. Carignan, C. Caloz, and D. Menard, 2010 IEEE MTT-S Int. Microw. Symp. 1336 (2010).

[35] L. Li and J. Cai, (2012).

[36] A. Hillion, A. Tamion, F. Tournus, C. Albin, and V. Dupuis, Phys. Rev. B **95**, 1 (2017).

[37] C. Dobrot and A. Stancu, Phys. B **457**, 280 (2015).

[38] A. Stancu, C. Pike, L. Stoleriu, P. Postolache, and D. Cimpoesu, J. Appl. Phys. **93**, 6620 (2003).

[39] C.I. Dobrotă and A. Stancu, J. Phys. Condens. Matter **25**, (2013).

[40] C.I. Dobrotă and A. Stancu, Phys. B Condens. Matter **407**, 4676 (2012).

[41] J. Geshev, J. Magn. Magn. Mater. **467**, 135 (2018).

[42] J. Geshev, L.L. Bianchi, R.F. Lopes, J.L. Salazar Cuaila, and A. Harres, J. Magn. Magn. Mater. **497**, 166061 (2019).

[43] O. Henkel, Phys. Status Solidi **7**, 919 (1964).

[44] P.E.Z. Ice, **25**, 3881 (1989).

[45] C. Blanco-Andujar, D. Ortega, P. Southern, Q.A. Pankhurst, and N.T.K. Thanh, Nanoscale **7**, 1768 (2015).

[46] T. Wang, Y. Wang, Y. Fu, T. Hasegawa, H. Oshima, K. Itoh, K. Nishio, H. Masuda, F.S. Li, H. Saito, and S. Ishio, Nanotechnology **19**, (2008).

[47] R.J. Veitch, IEEE Trans. Magn. **26**, 1876 (1990).

[48] F. Beron, L. Clime, M. Ciureanu, D. Menard, R.W. Cochrane, and A. Yelon, J. Nanosci. Nanotechnol. **8**, 2944 (2008).

[49] J.M. Martnez Huerta, J. De La Torre Medina, L. Piraux, and A. Encinas, J. Appl. Phys. **111**, (2012).

[50] A. Ramazani, V. Asgari, A.H. Montazer, and M.A. Kashi, Curr. Appl. Phys. **15**, 819 (2015).

[51] M.R. Zamani Kouhpanji and B.J.H. Stadler, ArXiv:1911.12480 1 (2019).

[52] C. Carvallo, A.R. Muxworthy, and D.J. Dunlop, Phys. Earth Planet. Inter. **154**, 308 (2006).

[53] A.P. Roberts, D. Heslop, X. Zhao, and C.R. Pike, Am. Geophys. Union **52**, 557 (2014).

[54] M.R. Zamani Kouhpanji, P.B. Visscher, and B.J.H. Stadler, Submitt. to ACS Nano (2020).

[55] M.R. Zamani Kouhpanji, A. Ghoreyshi, and B.J.H. Stadler, Sci. Rep. (2020).

[56] P.B. Visscher, AIP Adv. **035117**, (2019).

[57] X. dong Che and H. Neal Bertram, J. Magn. Magn. Mater. **116**, 121 (1992).

[58] M. Rivas, P. Gorria, C. Muñoz-Gómez, and J.C. Martińez-García, IEEE Trans. Magn. **53**, 2 (2017).

[59] F. Groß, S.E. Ilse, G. Schütz, J. Gräfe, and E. Goering, Phys. Rev. B **99**, (2019).

[60] F. Beron, L. Clime, M. Ciureanu, D. Menard, R.W. Cochrane, and A. Yelon, J. Appl. Phys. **101**, (2007).

[61] C. Pike and A. Fernandez, J. Appl. Phys. **85**, 6668 (1999).

[62] R.J. Harrison and J. Feinberg, Geochemistry Geophys. Geosystems **9**, (2008).

[63] D. Cimpoesu, I. Dumitru, and A. Stancu, J. Appl. Phys. **125**, (2019).

[64] A. Rotaru, J. Lim, D. Lenormand, A. Diaconu, J.B. Wiley, P. Postolache, A. Stancu, and L. Spinu, Phys. Rev. B **134431**, 1 (2011).

[65] F. Béron, L. Clime, M. Ciureanu, D. Ménard, R.W. Cochrane, A. Yelon, and L. Fellow, IEEE Trans. Magn. **42**, 3060 (2006).